\documentclass[graybox]{svmult}

\usepackage{mathptmx}       
\usepackage{helvet}         
\usepackage{courier}        
\usepackage{type1cm}        
\usepackage{makeidx}         
\usepackage{graphicx}        
\usepackage{multicol}        
\usepackage[bottom]{footmisc}

\makeindex             

\begin{document}

\title*{Einstein-Charged Scalar Field Theory: Black Hole
Solutions and Their Stability}
\author{Supakchai Ponglertsakul, Sam Dolan and Elizabeth Winstanley}
\institute{Supakchai Ponglertsakul \at School of Mathematics and Statistics, The University of Sheffield, Hicks Building, Hounsfield
Road, Sheffield, S3 7RH, United Kingdom \email{smp12sp@sheffield.ac.uk}
\and Sam Dolan \at School of Mathematics and Statistics,
The University of Sheffield, Hicks Building, Hounsfield Road,
Sheffield, S3 7RH, United Kingdom \email{S.Dolan@sheffield.ac.uk}
\and Elizabeth Winstanley \at School of Mathematics and Statistics,
The University of Sheffield, Hicks Building, Hounsfield Road,
Sheffield, S3 7RH, United Kingdom, \email{E.Winstanley@sheffield.ac.uk}}
%
%
\maketitle

\abstract{A complex scalar field on a charged black hole in a cavity is known to experience a superradiant instability. We investigate possible final states of this instability. We find hairy black hole solutions of a fully coupled system of Einstein gravity and a charged scalar field. The black holes are surrounded by a reflecting mirror. We also investigate the stability of these black holes.}

\section{Introduction}\label{sp:sec1}
In black hole physics, there is a mechanism where rotational (electromagnetic) energy can be extracted from a rotating (charged) black hole. This is called superradiant scattering. More specifically, the amplitude of a scalar field around a  black hole will be amplified if the frequency $\sigma$ of the field satisfies \cite{Bekenstein:1973mi} $\sigma < m \varOmega_{\mathrm{H}} + q \varPhi_{\mathrm{H}}$, where $m, q, \varOmega_{\mathrm{H}}$ and $\varPhi_{\mathrm{H}}$ are the  azimuthal quantum number, scalar field charge, angular velocity and electric potential at the outer horizon respectively. By setting $\varOmega_{\mathrm{H}}=0$, the charged version of the superradiant condition is obtained.

One can create an instability of the spacetime background via a superradiant scattering process, if there is some mechanism to confine the bosonic field within the vicinity of the black hole. Then wave modes will be repeatedly scattered off the black hole and their amplitude will be intensified. The back-reaction of the field modes on the background will eventually become significant. In the charged case, the trapping mechanism can be induced by either (i) a reflecting mirror \cite{Herdeiro:2013} or (ii) anti-de Sitter boundary condition \cite{Uchikata:2011zz}. By having either of these with a charged black hole, the superradiant instability can be triggered.

An interesting question that one might ask is, what is the end-point of this charged-scalar superradiant instability? To tackle this problem, a fully non-linear analysis is required. Hence in this talk, we investigate possible end-points of the superradiant instability for charged black holes with a reflecting mirror. More specifically, a coupled system of gravity and a massless complex scalar field with a mirror-like boundary condition is studied. Numerical black hole solutions with a non-trivial scalar field are obtained. By considering linearised perturbations of these black holes, a numerical analysis of the black hole's stability is undertaken. Here we present a selection of plots to illustrate our numerical results. More details of this work can be found in Ref.~\cite{SP:2015}.


\section{Linear Perturbations in Electrovacuum}\label{sp:sec2}
In this section, a massless complex scalar field $\phi$ on the Reissner-Nordstr\"{o}m (RN) background in a cavity is considered. In the test-field limit, the dynamics of a scalar field on RN spacetime is described by the Klein-Gordon (KG) equation. By substituting the ansatz $\phi\sim e^{-i\sigma t}R(r)$, where $\sigma$ and $R(r)$ are respectively the frequency and the radial part of the scalar field, into KG equation, we obtain a second order differential equation in terms of the radial part. To solve this equation, we apply the following boundary conditions: (i) an ingoing wave near the horizon $r\rightarrow r_{\mathrm{h}}$ and (ii) at the mirror the scalar field vanishes, so $R(r_{\mathrm{m}})=0$, where $r_{\mathrm{m}}$ is the location of the mirror. Then a numerical technique called the shooting method is implemented. We scan for corresponding frequencies $\sigma$ such that the perturbations satisfy the boundary conditions.

\begin{figure}
\begin{center}
\begin{tabular}{cc}
  \includegraphics[width=2.2in]{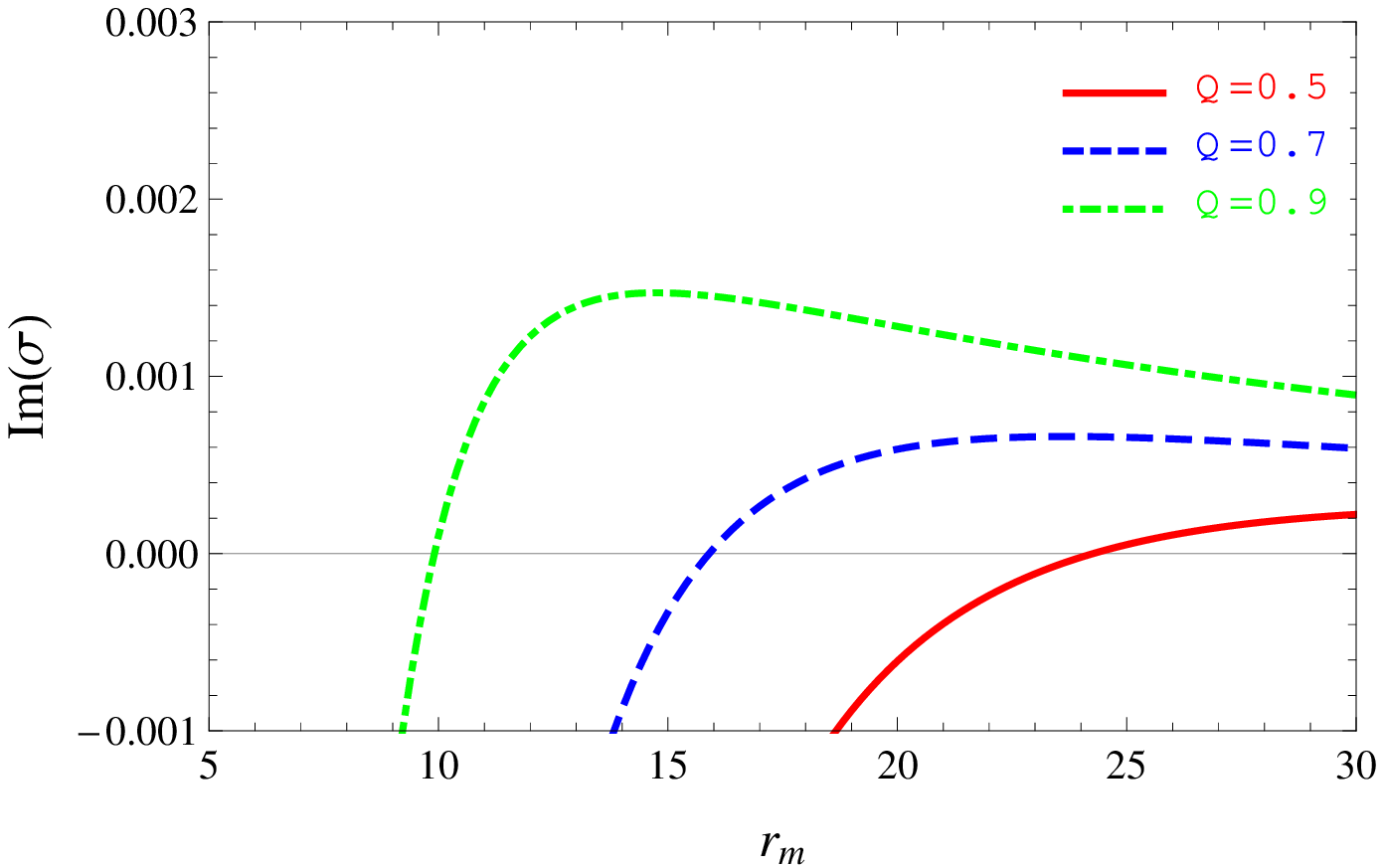} &
  \includegraphics[width=2.2in]{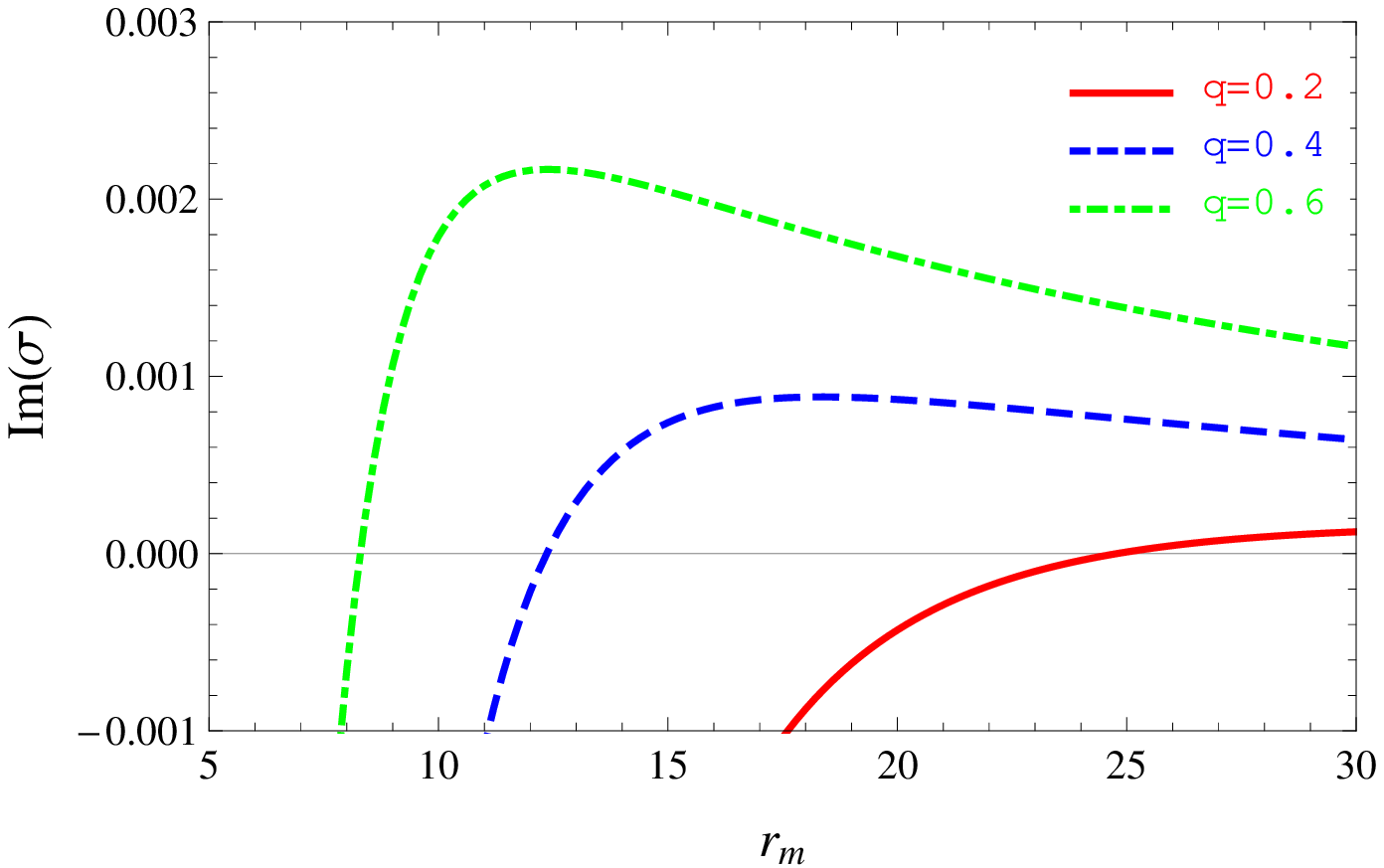}  \\
  \footnotesize{(a)} &
    \footnotesize{(b)}
    \end{tabular}
    \end{center}
  \caption{The imaginary part of $\sigma$ is plotted as a function of the location of the mirror $r_{\mathrm{m}}$, (a) for fixed scalar charge $q=0.5$ and different values of the black hole charge $Q$, (b) for fixed $Q=0.9$ and different values of $q$. Taken from \cite{SP:2015}}
  \label{sp:fig1}
\end{figure}

With the black hole mass fixed to be $M=1$, the example plot below (Fig.~\ref{sp:fig1}) illustrates the frequency as a function of the location of mirror. It is clear from Fig.~\ref{sp:fig1}(a) and \ref{sp:fig1}(b) that this system experiences a superradiant instability as there are regions where Im$(\sigma)>0$, indicating an unstable mode. One can learn the following from these plots: when the location of mirror $r_{\mathrm{m}}$ is small (close to black hole), the field mode decays exponentially in time; instability occurs when $r_{\mathrm{m}}$ reaches a certain value.

We find that a massless charged scalar field on the RN background in a cavity experiences a superradiant instability. These results are in agreement with with previous work done by Herdeiro et al. \cite{Herdeiro:2013}, where they studied a massive complex scalar field on a charged black hole with a mirror. To fully understand what could happen at the end-point of this instability, a non-linear system of gravity and a charged scalar field must be investigated. In the next section, a fully coupled Einstein-charged scalar system will be considered.

\section{Static Black Holes}\label{sp:sec3}
The self-gravitating system of a charged scalar field is described by the following action
\begin{equation}\label{sp:eq2}
S  = \int d^{4}x \sqrt{-g} \left[ \frac{R}{16\pi G} - \frac{1}{4} F_{ab}F^{ab} -\frac{1}{2}g^{ab} D^\ast_{(a} \phi^\ast D^{}_{b)} \phi \right].
\end{equation}
In Eq.~(\ref{sp:eq2}) Faraday tensor is defined by $F_{ab} = \nabla_a A_b - \nabla_b A_a$, where $A_a$ is the electromagnetic potential, $D_a = \nabla_a - iqA_a$, $q$ is the scalar field charge and $X_{(ab)}=\frac{1}{2}\left(X_{ab}+X_{ba}\right)$.
 Varying Eq.~(\ref{sp:eq2}), we obtain three equations of motion
\begin{eqnarray}
G_{ab} &=& 8\pi G \left(T_{ab}^F + T_{ab}^{\phi}\right),\label{sp:eq3} \\
\nabla_a F^{ab} &=& \frac{iq}{2} \left(\phi^\ast D^b \phi - \phi (D^b \phi)^\ast \right),\label{sp:eq4} \\
D_a D^a \phi &=& 0. \label{sp:eq5}
\end{eqnarray}
We consider a static spherically symmetry black hole spacetime with line element
\begin{equation}
ds^2 = -f(r)h(r)dt^2 + f(r)^{-1}dr^2 + r^2\left(d\theta^2 + \sin^2\theta~d\varphi^2\right).\label{sp:eq6}
\end{equation}
In addition, the electromagnetic vector potential is $A_a\equiv[A_{0}(r),0,0,0]$ and the scalar field depends on $r$ only $\phi=\phi(r)$. By inserting these ansatzes into Eqs.~(\ref{sp:eq3}--\ref{sp:eq5}), we obtain three coupled ordinary differential equations. By imposing appropriate boundary conditions at the event horizon and at the mirror, these coupled equations can be solved numerically. We also require that the scalar field must vanish at the location of the mirror.

\begin{figure}
\begin{center}
\begin{tabular}{cc}
  \includegraphics[width=2.2in]{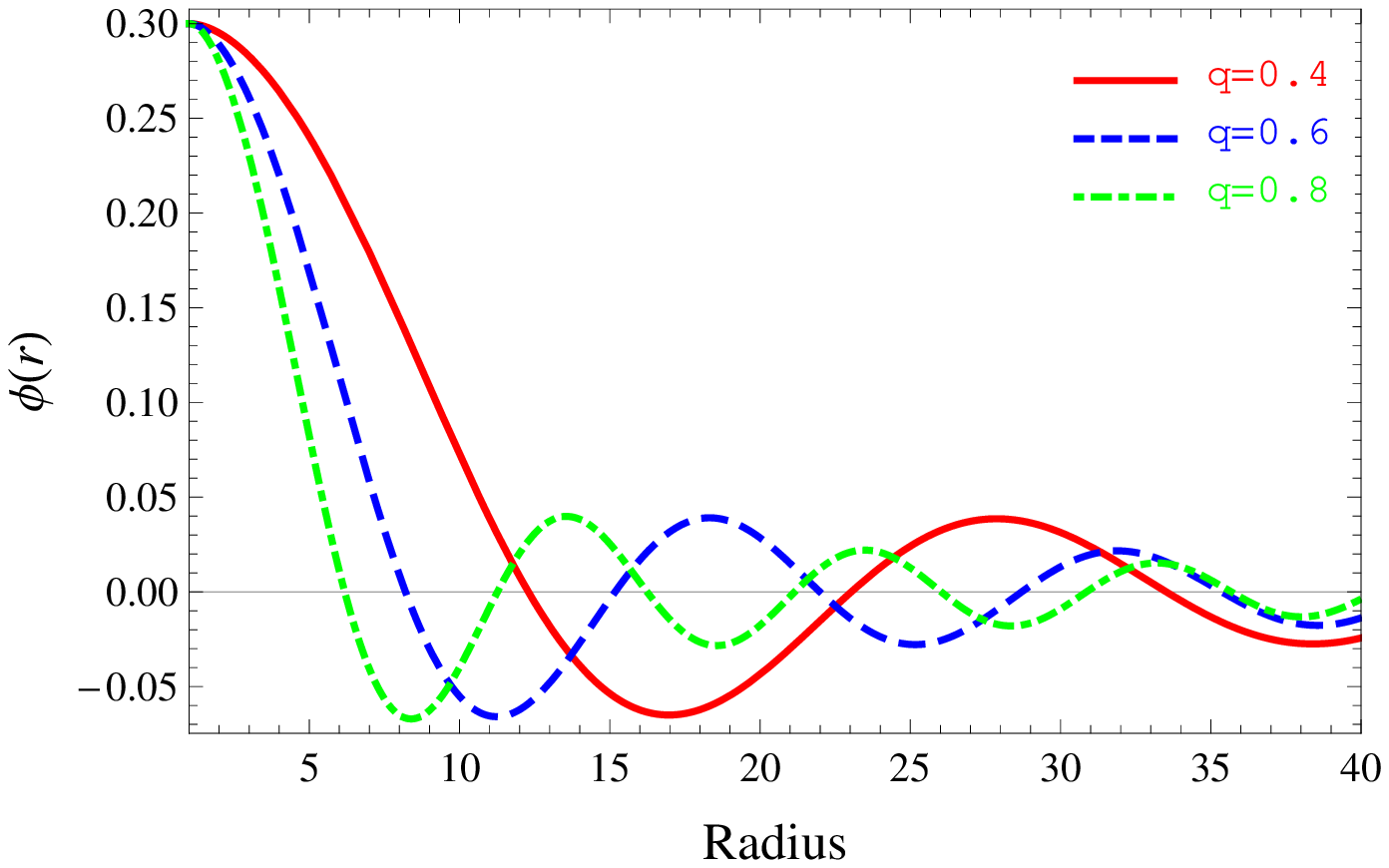} &
  \includegraphics[width=2.2in]{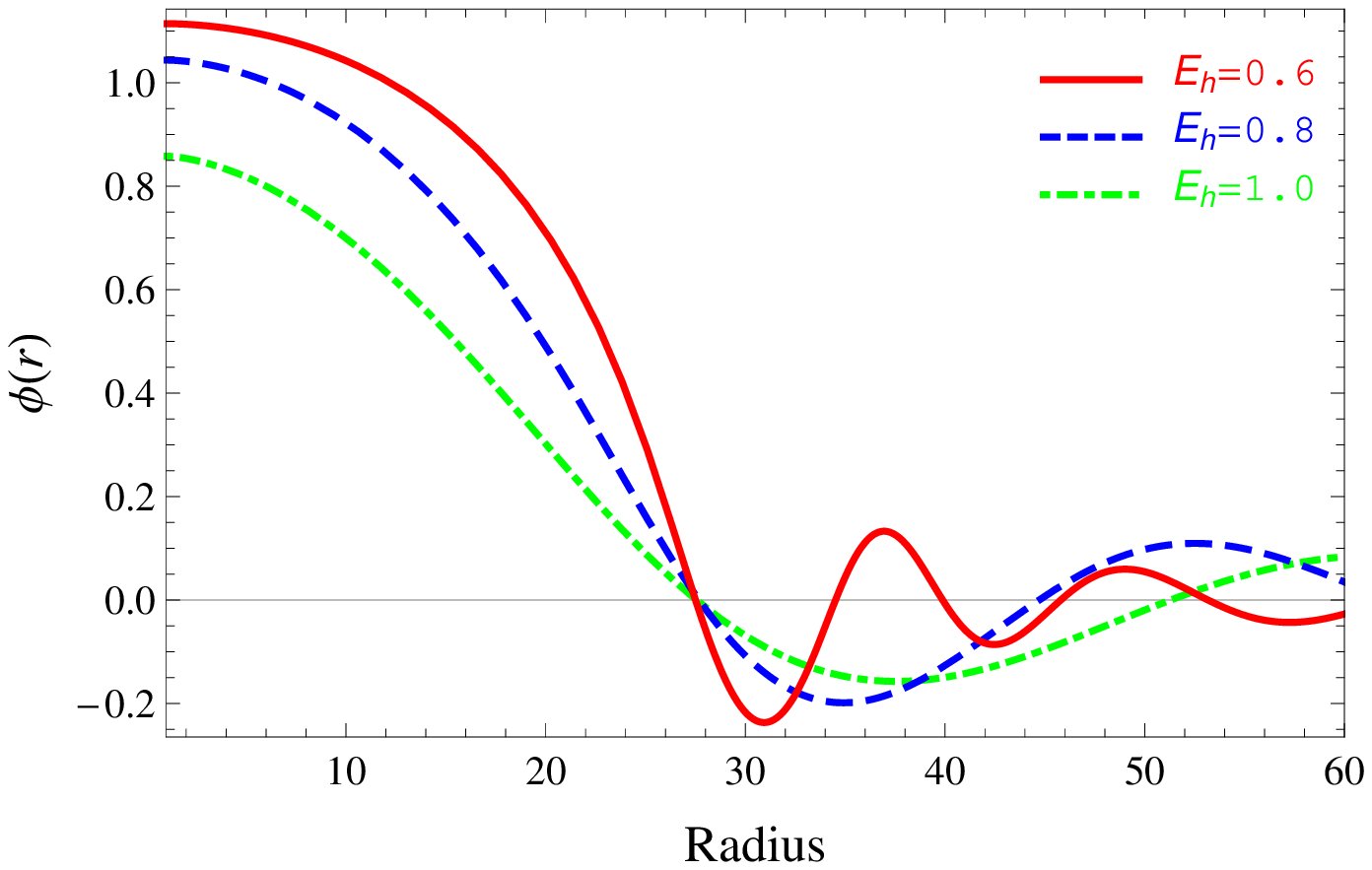} \\
  \footnotesize{(a)} & \footnotesize{(b)}
  \end{tabular}
  \end{center}
  \caption{The scalar field $\phi(r)$ is plotted as a function of radius (a) for fixed $\phi_{\mathrm{h}}=0.3$, $E_{\mathrm{h}}=0.6$ and different values of $q$, (b) for fixed $q=0.1$ and different values of $E_h$. Taken from \cite{SP:2015}}%
 \label{sp:fig2}
\end{figure}

To obtain static solutions, three parameters must be specified: $\phi_{\mathrm{h}},$ the value of the scalar field on the horizon; $E_{\mathrm{h}} \equiv A'_{0}(r_{\mathrm{h}}),$ the electric field on the horizon; and $q$. In Fig.~\ref{sp:fig2}, we show some example solutions where the black hole radius is fixed at $r_h = 1$. The nontrivial structure of the scalar field can be seen. Because these scalar fields oscillate around zero therefore, one can put the mirror at any zero of $\phi$, however, in this work, we only consider the case where the mirror is located at the first zero, since these solutions are expected to be stable.

By varying the three parameters $\phi_{\mathrm{h}}, E_{\mathrm{h}}$ and $q$, we obtain different hairy black hole solutions. In Fig.~\ref{sp:fig2}(a), three distinct black hole solutions with three different scalar charges are displayed. Note that these solutions possess three different mirror radii. However, it is possible that different static solutions can share the same mirror location as illustrated in Fig.~\ref{sp:fig2}(b).

\section{Stability of the Hairy Black Holes}\label{sp:sec4}
In the previous section, we have shown that Einstein-charged scalar field theory in a cavity allows the existence of hairy black holes. Our next important question is, are these solutions stable or unstable? If they are shown to be stable, they could represent a possible end-point of the superradiant instability for a massless charged scalar perturbation on the RN background with a mirror.

We consider linear spherically symmetric perturbations, where the four field variables $(f,h,A_{0},\phi)$ are rewritten as follows, $f=\bar{f}(r) + \delta f(t,r)$ and similarly for the other three quantities. In this notation, $\bar{f}$ is the equilibrium quantity and $\delta f$ is the perturbed part. By linearising the field equations, we arrive at three coupled perturbation equations in terms of $\delta A_0$ and the real and imaginary parts of $\delta \phi$. These perturbation equations are very complicated and lengthy, full details of these equations can be found in \cite{SP:2015}. The three perturbation equations consist of two dynamical equations describing the real and imaginary parts of the scalar field respectively, the other one is a constraint equation. By substituting the ansatz $\delta\phi\sim e^{-i\sigma t}\tilde{\phi}(r)$ (and similarly for the other perturbations) into the equations, we can integrate the perturbation equations numerically. The perturbation modes are required to satisfy the boundary conditions that $\tilde{\phi}(r)$ and other perturbation modes have an ingoing wave-like condition near the horizon, and at the mirror the scalar field perturbations must vanish $(\tilde{\phi}(r_{\mathrm{m}})=0)$.

\begin{figure}
\begin{center}
\begin{tabular}{cc}
 \includegraphics[width=2.2in]{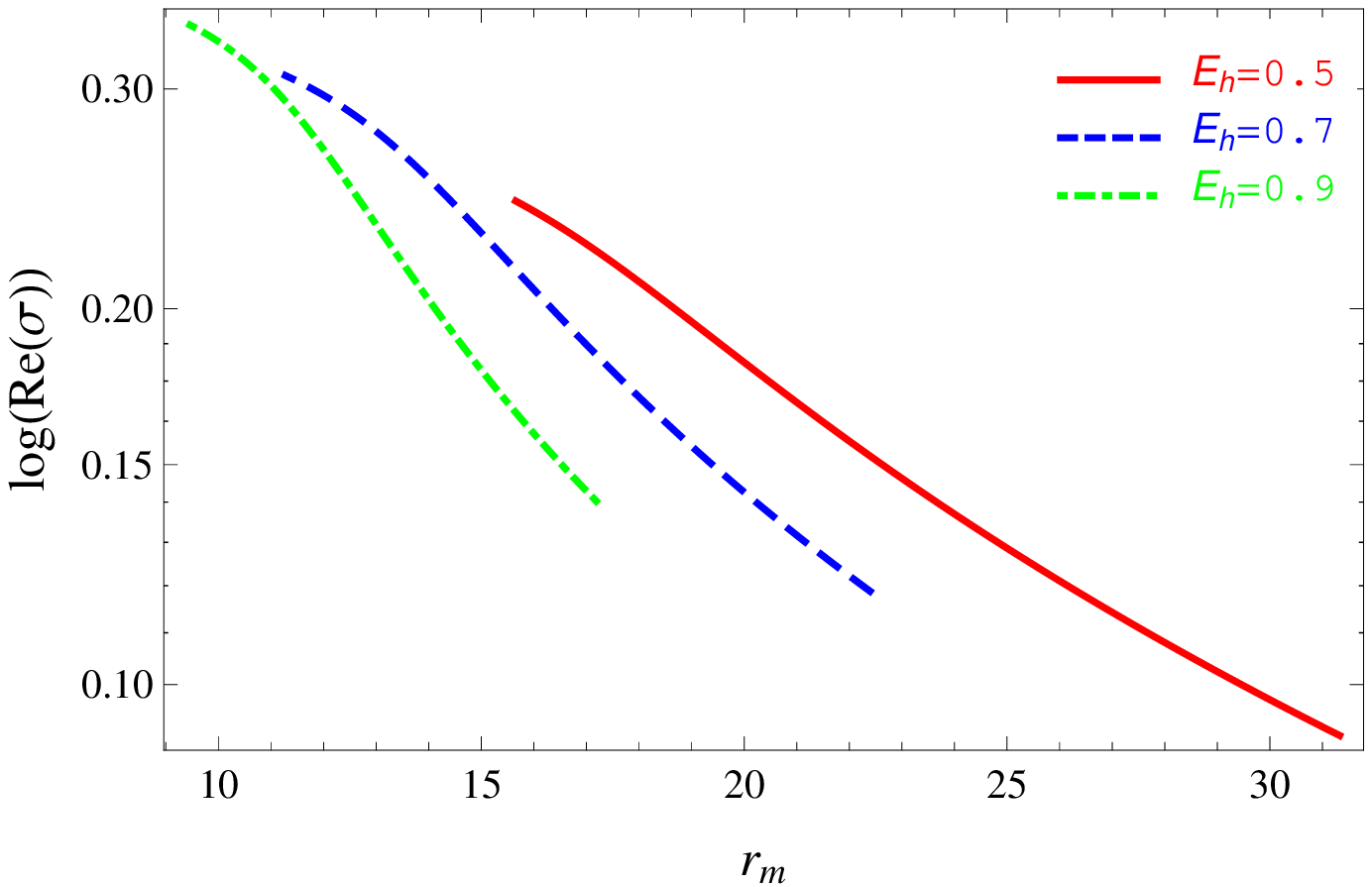} & 
  \includegraphics[width=2.2in]{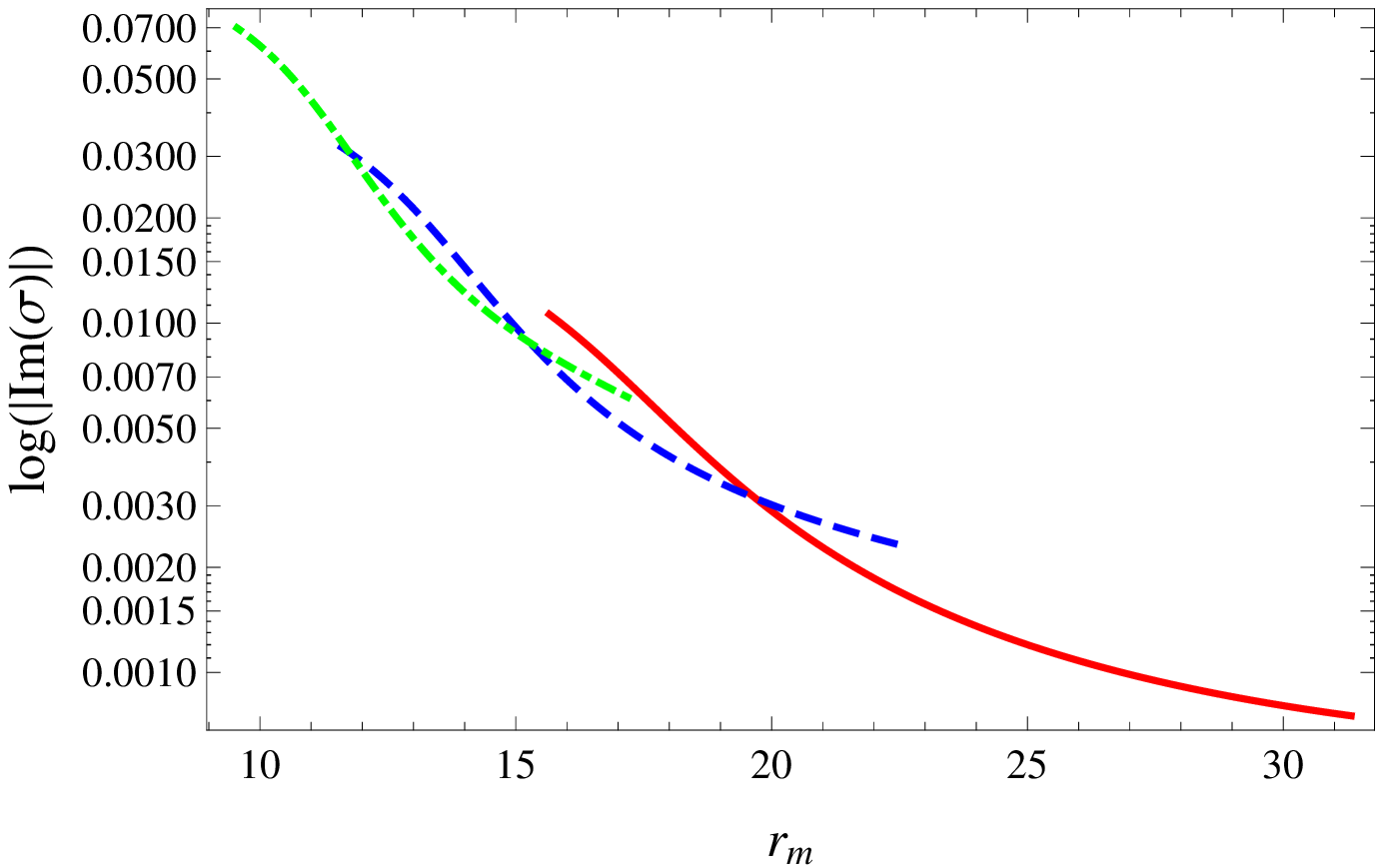}  \\ 
  \footnotesize{(a)} & \footnotesize{(b)}
  \end{tabular}
  \end{center}
  \caption{The real (a) and imaginary parts (b) of the perturbation frequency $\sigma$ are plotted against the location of the mirror $r_{\mathrm{m}}$ with the scalar charge fixed to be $q=0.2$, $\phi_{\mathrm{h}}$ varying from $0.1$ to $1.3$ and different values of $E_h$. We find that Im$(\sigma)<0$. Taken from \cite{SP:2015}}
    \label{sp:fig3}
\end{figure}

\begin{figure}
\begin{center}
\begin{tabular}{cc}
  \includegraphics[width=2.2in]{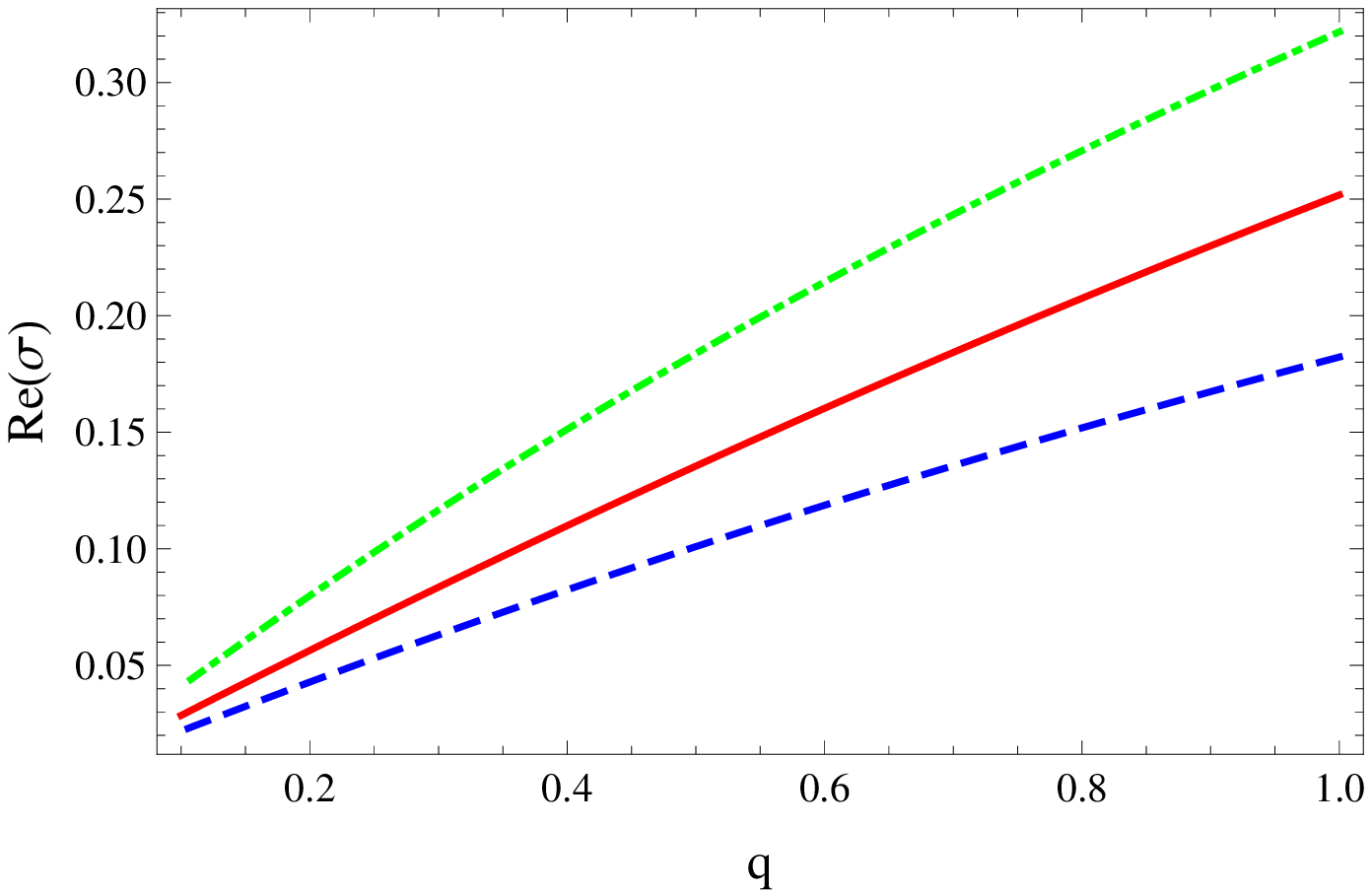} &
  \includegraphics[width=2.2in]{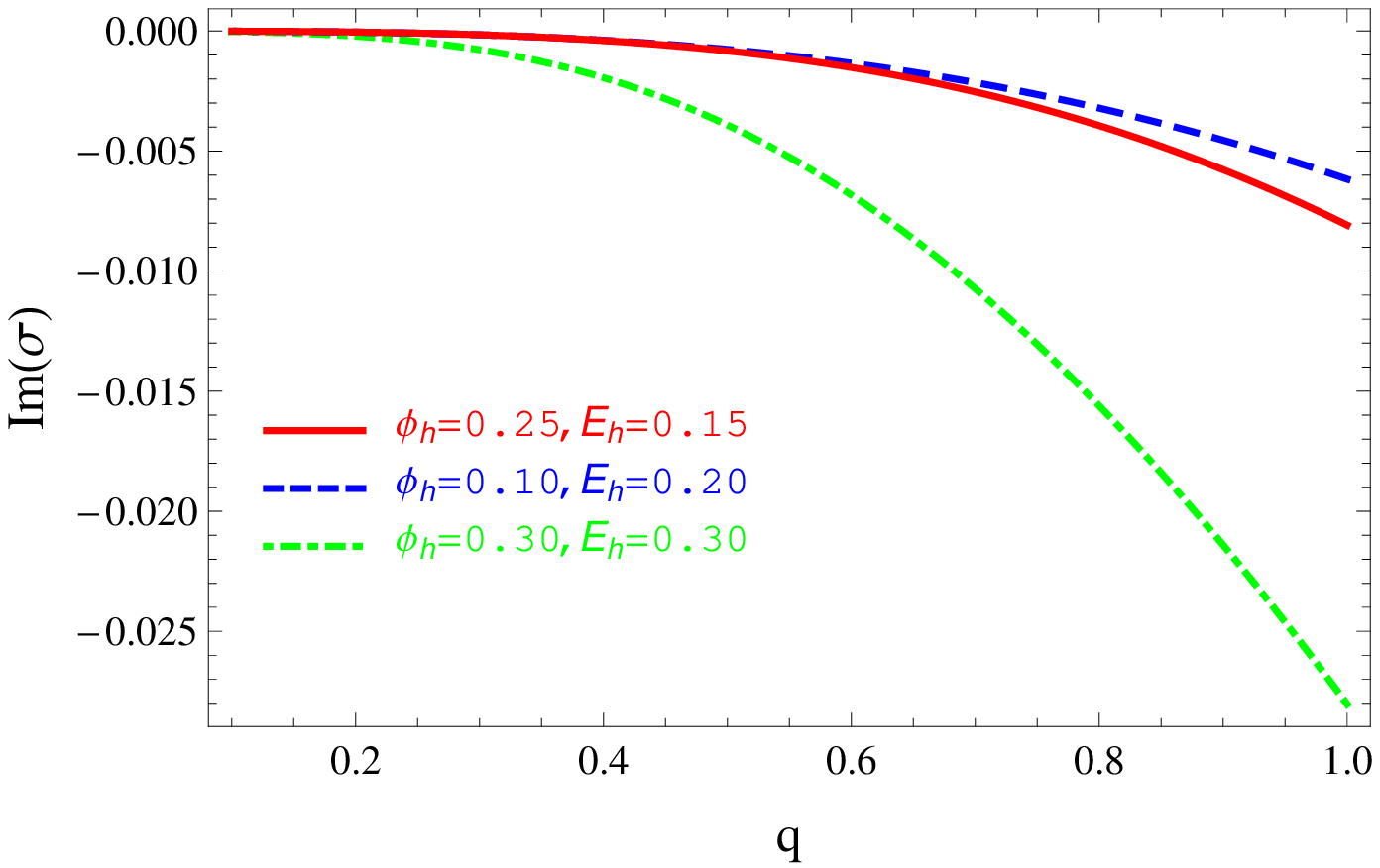} \\
\footnotesize{(a)} & \footnotesize{(b)}
\end{tabular}
\end{center}
  \caption{The real (a) and imaginary parts (b) of the perturbation frequency $\sigma$ are plotted against the scalar charge $q$ with a selection of various values of $\phi_{\mathrm{h}}$ and $E_{\mathrm{h}}$. Taken from \cite{SP:2015}}
  \label{sp:fig4}
\end{figure}

The numerical scheme is as follows. Firstly, static background parameters are specified, $q,\phi_{\mathrm{h}}$ and $E_{\mathrm{h}}$, then the equilibrium field equations are integrated. From the solution, we locate the first zero of  the equilibrium scalar field, setting this to be the location of the mirror $r_{\mathrm{m}}$. Then the coupled perturbation equations are solved by scanning for frequencies $\sigma$ such that the perturbations satisfy the required boundary conditions.

In Fig.~\ref{sp:fig3}, the real and imaginary parts of $\sigma$ are plotted as functions of the mirror radius $r_m$ for scalar field charge $q=0.2$. In each plot, the parameter describing static solution varies between $\phi_{\mathrm{h}}=0.1-1.3$. Thus each point in this plot refers to the perturbation frequency for one distinct hairy black hole. With the mirror at the first zero of the static scalar field $\phi$, we find one value of $\sigma$ for which the perturbations satisfy the boundary conditions. Fig.~\ref{sp:fig3}(b) shows that the perturbation modes decay exponentially in time since we find that Im$(\sigma)<0$. In addition, the real and imaginary parts of the perturbation frequency $\sigma$ are plotted as functions of the scalar charge $q$ are displayed in Fig.~\ref{sp:fig4}. Here in this example, for each curve, a selection of values of static black hole parameters $\phi_{\mathrm{h}}$ and $E_{\mathrm{h}}$ are fixed. Fig.~\ref{sp:fig4}(b) illustrates that the perturbation modes are exponentially decaying in time. We refer the reader to \cite{SP:2015} where we find Im$(\sigma)<0$ for all black hole solutions investigated.

\section{Summary}\label{sp:sec5}
We have studied a coupled system involving Einstein gravity and a complex charged scalar field in the presence of a mirror-like boundary condition. Numerical hairy black hole solutions were obtained by the shooting method. By putting a mirror at the first node of the equilibrium scalar field, we showed these black hole solutions are stable under spherically symmetric linear perturbations. Therefore, we conclude that these black holes could represent an end-point of the superradiant instability of Reissner-Nordst\"{o}rm black holes to charged scalar field perturbations.

\begin{acknowledgement}
The work of SRD and EW is supported by the Lancaster-Manchester-Sheffield Consortium for Fundamental Physics under STFC grant ST/L000520/1. The work of SRD is also supported by EPSRC grant EP/M025802/1. The work of SP is supported by the 90th Anniversary of Chulalongkorn University Fund (Ratchadaphiseksomphot Endowment Fund).
\end{acknowledgement}
%
%


\end{document}